\documentclass[prb,aps,twocolumn,floatfix,showpacs, showkeys,]{revtex4-1}

\usepackage{graphicx}
\usepackage{epstopdf}
\usepackage{upgreek}

\usepackage{color}

\usepackage[utf8]{inputenc}

\begin{document}
\title{Magnetism and charge density waves in RNiC$_2$ (R = Ce, Pr, Nd)}

\author{Kamil K. Kolincio}
\email{kkolincio@mif.pg.gda.pl}
\affiliation{Faculty of Applied Physics and Mathematics, Gdansk University of Technology,
Narutowicza 11/12, 80-233 Gdansk, Poland}
\author{Marta Roman, Micha\l{} J. Winiarski, Judyta Strychalska - Nowak}
\affiliation{Faculty of Applied Physics and Mathematics, Gdansk University of Technology,
Narutowicza 11/12, 80-233 Gdansk, Poland}
\author{Tomasz Klimczuk} 
\email{tomasz.klimczuk@pg.edu.pl}
\affiliation{Faculty of Applied Physics and Mathematics, Gdansk University of Technology,
Narutowicza 11/12, 80-233 Gdansk, Poland}

\begin{abstract}
We have compared the magnetic, transport, galvanomagnetic and specific heat properties of CeNiC$_2$, PrNiC$_2$ and NdNiC$_2$ to study the interplay between charge density waves and magnetism in these compounds. The negative magnetoresistance in NdNiC$_2$ is discussed in terms of the partial destruction of charge density waves and  an irreversible phase transition stabilized by the field induced ferromagnetic transformation is reported. For PrNiC$_2$ we demonstrate that the magnetic field initially weakens the CDW state, due to the Zeeman splitting of conduction bands. However, the Fermi surface nesting is enhanced at a temperature related to the magnetic anomaly.
\end{abstract}
\maketitle

\section{introduction}
The interaction between charge density waves (CDW) and different types of orderings such as superconductivity \cite{Chang2012, daSilva2014, Chang2016}, spin density waves \cite{Fawcett1988, Jacques2014, Young1974} and magnetism\cite{Balseiro1980} has been a long standing area of interest. Magnetic order or applied magnetic field have been found to impact the CDW state through changing the geometry of the Fermi surface (FS). The effect can be destructive due to the disturbance of the FS nesting caused by the magnetic field-induced splitting of the conduction bands or modification of the electronic structure due to a magnetic transition \cite{Tiedje1975}. Alternatively, a constructive effect has been observed in a group of materials, in which this FS transformation leads to the enhancement of the nesting conditions or when the nesting vector has the ability to adapt to the evolution of the Fermi surface\cite{Brooks2006, Wang2014, Andres2011, Zanchi1996, Graf2004, Winter2013, Murata2015}. Recently, much attention of the researchers exploring the coupling between CDW, superconductivity and magnetic order has been devoted to the two families of ternary compounds: M$_5$Ir$_4$Si$_{10}$, (where M = Y, Dy, Ho, Er, Tm, Yb or Lu )\cite{Lalngilneia2015, vanSmaalen2004, Galli2000, Galli2002, Hossain2005, Leroux2013, Singh2004, Sangeetha2012, Kuo2006} and RNiC$_2$, (where R = La, Ce, Pr, Nd, Sm, Gd or Tb)\cite{Kim2013, Prathiba2016}. Most of the members of the latter family exhibit the Peierls transitions towards the charge density wave state \cite{murase2004}. The relevance of a Peierls instability has been confirmed for R = Gd, Tb, Nd, Pr and Sm, while the LaNiC$_2$ and CeNiC$_2$ compounds do not show any anomalies that could be attributed to CDW \cite{Yamamoto2013, Laverock2009, ahmad2015, Shimomura2016, Wolfel2010}. Instead, LaNiC$_2$ is an unconventional noncentrosymmetric superconductor with $T_c$ = 2.7 K \cite{Lee1996,Pecharsky1998,Wiendlocha2016}. Next to the CDW, the members of the RNiC$_2$ family show a wide range of magnetic orderings originating from the RKKY interaction between local magnetic moments and conduction electrons \cite{Schafer1997, Kotsanidis1989}. The ground state of RNiC$_2$ depends on the rare-earth atom marked in the above formula by R: CeNiC$_2$, NdNiC$_2$, GdNiC$_2$ and TbNiC$_2$ show the antiferromagnetic character \cite{Bhattacharyya2014, Yakinthos1990, Hanasaki2011, Uchida1995, matsuo1996, Pecharsky1998}, SmNiC$_2$ is a ferromagnet, while the PrNiC$_2$ compound has been identified as a van Vleck paramagnet \cite{Onodera1998}. This rich variety of the types of magnetic ordering shown by the RNiC$_2$ family members motivated us to explore the interplay of charge density waves and various magnetic ground states. Here, we compare the physical properties of three isostructural, yet highly dissimilar compounds: NdNiC$_2$, PrNiC$_2$ and CeNiC$_2$. The first compound, NdNiC$_2$ shows the Peierls instability with $T_P$ = 121 K and antiferromagnetic ordering with $T_N$ = 17 K. The second, PrNiC$_2$ undergoes the CDW transition at $T_P$ = 89 K and instead of long range magnetic ordering, shows a magnetic anomaly at $T^*$ = 8 K. The last compound, CeNiC$_2$ becomes an antiferromagnet at $T_N$ = 20 K and does not exhibit the CDW transition. 
\section{experimental details}
The polycrystalline samples of RNiC$_2$ (where R = Ce, Pr, and Nd) were synthesized by arc-melting the stoichiometric amounts of pure elements: Ni (4N), C (5N) and Ce (3N), Pr (3N), Nd (3N) in a high purity argon atmosphere. Small excess of Ce, Pr, Nd ($\approx2\%$) and C ($\approx5\%$) was used to compensate the loss during arc-melting. To obtain good homogeneity of samples, the specimens were turned over and remelted four times in a water-cooled copper hearth. A zirconium button was used as an oxygen getter. The buttons obtained from the arc-melting process were wrapped in tantalum foil, placed in evacuated quartz tubes, annealed at 900$^o$C for 12 days and cooled down to the room temperature by quenching in cold water. Overall mass loss after the melting and annealing processes were negligible ($\approx1\%$). 

The low temperature experiments were performed with a Quantum Design physical properties measurements system (PPMS) allowing for the application of a magnetic field as large as 9 T.
Thin Pt wires ($\phi$ = 37 $\mu$m) serving as electrical contacts for transport and Hall measurements were spark-welded to the polished sample surface. A standard four-probe contact configuration was used to measure resistivity. A magnetic field was applied perpendicularly to the current direction.
The Hall voltage was collected in reversal directions of magnetic field in order to remove the parasitic longitudinal magnetoresistance voltage due to misalignment of electrical contacts.
The specific heat measurements were performed using the dual slope method on flat polished samples.
 Magnetization measurements were carried out using the ACMS susceptometry option of the PPMS system. Pieces of the samples were fixed in standard polyethylene straw holders.

\section{Results and discussion}
The phase composition and crystallographic structure of the samples were checked by powder X-ray diffraction (pXRD) at room temperature. The pXRD analysis shows that all observed peaks for NdNiC$_2$ and PrNiC$_2$ are successfully indexed in the orthorhombic CeNiC$_2$-type structure\cite{matsuo1996} with a space group Amm2 (\# 38), which confirms the phase purity of the obtained samples. Only for the CeNiC$_2$ sample, additional reflections corresponding to a small amount of the secondary phase\cite{Sakai1980} CeC$_2$ are observed. The lattice parameters were determined from the LeBail profile refinements of the pXRD patterns carried out using FULLPROF software\cite{FULLPROF}. The obtained values of the lattice constants, shown in Table \ref{parameters}  are in good agreement with those reported in the literature \cite{Motoya1997, Yakinthos1990, Onodera1998, Schafer1992}.
\begin{table}[ht]
 \caption{\label{parameters}Lattice constants, unit cell volume and the parameters of the LeBail refinements for CeNiC$_2$, PrNiC$_2$ and  NdNiC$_2$, at room temperature.}
 \begin{ruledtabular}\begin{tabular}{cccc}

 & CeNiC$_2$ & PrNiC$_2$ & NdNiC$_2$  \\
\hline 
a (\AA) & 3.8753(2)	& 3.8239(5) &	3.7834(1) \\
b (\AA) & 4.5477(2)	& 4.5428(8)	& 4.5361(1) \\
c (\AA) & 6.1601(3)	& 6.1448(1)	& 6.1285(1) \\
V (\AA$^3$)	& 108.565(8) &	106.746(3) &	105.178(3) \\
R$_p$ &	12.3 &	7.51 &	8.35 \\
R$_{wp}$ &	16.5 &	10.1 &	10.8 \\
R$_{exp}$ &	11.49 &	7.54 &	7.7 \\
$\chi^2$ &	2.05 &	1.81 &	1.96 \\

\end{tabular}\end{ruledtabular}
\end{table}

\begin{figure}[h!t]
 \includegraphics[scale=0.25]{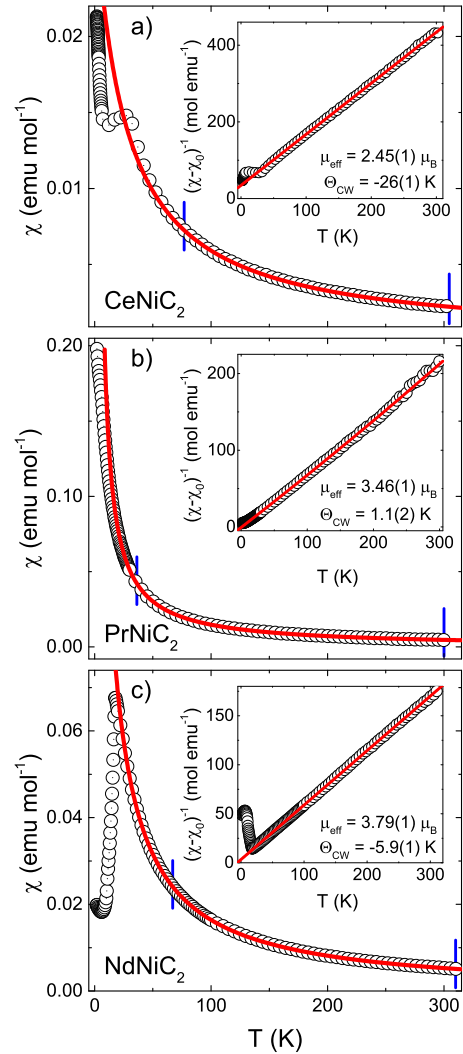}
  \caption{\label{MvsT_all} Magnetic susceptibility of CeNiC$_2$ (a), PrNiC$_2$ (b), and NdNiC$_2$ (c) at applied magnetic field $\mu_0 H$ = 1 T (open circles). Red lines show fits using the modified Curie-Weiss expression (Eq. \ref{CurieWeiss}). Insets show inverse susceptibilities displaying linear temperature dependence in agreement with the Curie-Weiss law (Eq. \ref{CurieWeiss}). Blue ticks mark the used fitting ranges. The effective magnetic moments extracted from fits agree with the values expected for free trivalent \textit{R} ions. Low-temperature part of susceptibility for PrNiC$_2$ is presented in Fig. \ref{Pr-Mdc_vs_T_PrNd_M_vs_H}}.
  \end{figure}
The temperature dependence of the magnetic susceptibility ($\chi$) measured at 1 T applied magnetic field is presented in Figure \ref{MvsT_all}. All three compounds show paramagnetic behavior at high temperatures. The $\chi(T)$ data were fitted using the modified Curie-Weiss expression:

\begin{equation}
\label{CurieWeiss}
\chi (T) = \frac{C}{T-\Theta_{CW}} + \chi_0
\end{equation}

 \begin{figure*}[h!t]
 \includegraphics[scale=0.2]{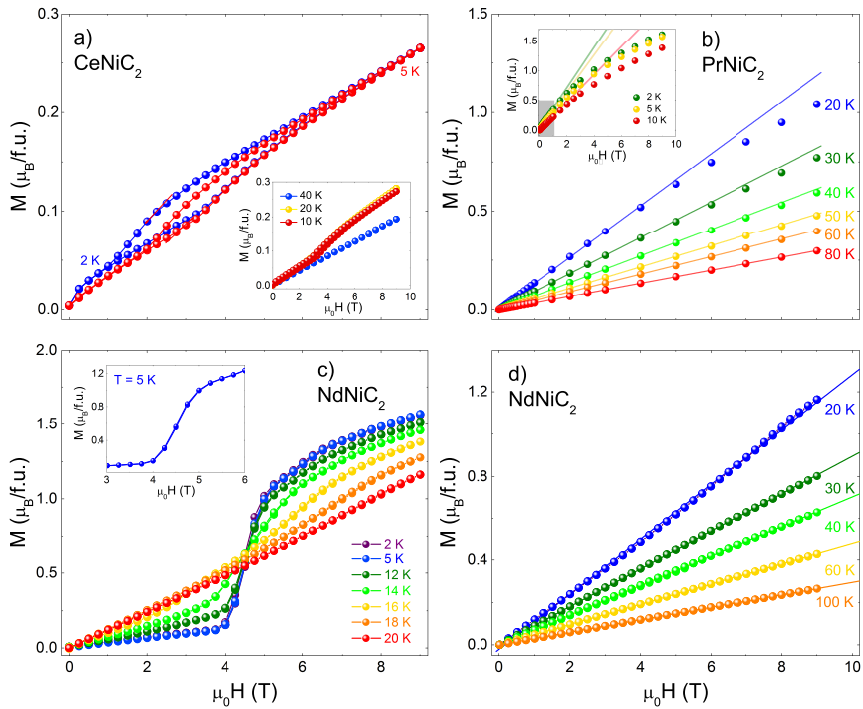}
  \caption{\label{Pr-Mdc_vs_T_PrNd_M_vs_H} Panel a) Magnetization vs. applied magnetic field ($M(H)$) measured for CeNiC$_2$ at 2 and 5 K (below the N\'{e}el temperature $T_N$ = 19 K) showing a hysteretic behavior probably due to a field-induced magnetic transition. The inset presents the magnetization at 10, 20, and 40 K. While the magnetization at $T\geq$ 40 K (above the AFM transition) is a linear function of applied field, in the vicinity (20 K) and below the $T_N$ an upturn is seen arround 3 T, suggesting the field-induced magnetic transition suppressing the AFM order. Panel b) presents $M(H)$ curves for PrNiC$_2$ showing linear character down to 40 K. Below that temperature the curves start to saturate in high magnetic fields. At the lowest temperatures (2, 5, and 10 K; see the inset) the deviation from linearity is clear above 1-2 T. Straight lines are least-squares linear fits to the low-field (below 1 T) magnetization data. Gray shading in the inset marks the fitting range used. Panel c) shows the low-temperature $M(H)$ data for NdNiC$_2$. At 20 K (above the $T_N$ = 17 K) the curve is linear up to 9 T while below this temperature an upturn is observed above approx. 4 T. In the temperatures lower than $T_N$ the magnetization below approx. 4 T is visibly suppressed due to AFM ordering of the magnetic moments. At 4 T a magnetic order-order transition results in rapid increase in magnetization. The inset shows magnetization around the field-induced magnetic transition at 5 K showing no sign of hysteresis. Panel d) presents magnetization of NdNiC$_2$ between 20 and 100 K, showing a linear character up to 9 T. Straight lines are least-squares linear fits to the low field data.}
  \end{figure*}
  
where $C$ is the Curie constant, $\Theta_{CW}$ is the Curie-Weiss temperature, and $\chi_0$ is the temperature-independent susceptibility resulting from both sample (Pauli and Van Vleck paramagnetism, Landau diamagnetism) and sample holder (small diamagnetic contribution of sample straw assembly). Having estimated the $C$ parameter and assuming that the magnetic moment originates from $R^{3+}$ ions only, one can calculate the effective magnetic moment using the relation shown in Equation \ref{EffectiveMoment}:

\begin{equation}
\label{EffectiveMoment}
\mu_{eff} = \sqrt{\frac{3 C k_B}{{\mu_B}^2 N_A}}
\end{equation}

where $k_B$ is the Boltzmann constant, $\mu_B$ is the Bohr magneton, and $N_A$ is Avogadro's number. The resulting effective magnetic moments of CeNiC$_2$, PrNiC$_2$ and NdNiC$_2$ are consistent with the values expected for free $R^{3+}$ ions\cite{Jensen1991}. The negative sign of $\Theta_{CW}$ obtained for the Ce- and Nd-bearing compounds (-26 K and -5.9 K, respectively) indicate an effectively antiferromagnetic coupling between the magnetic moments. In the case of PrNiC$_2$, the absolute value of $\Theta_{CW}$ is close to 0 suggesting the weakness or absence of magnetic interactions down to 2 K.

It is worth noting that the measured susceptibility of PrNiC$_2$ is well reproduced by the modified Curie-Weiss equation, yielding reasonable values of $C$, $\Theta_{CW}$, and $\chi_0$ and suggesting that the contribution of Pr$^{3+}$ local moments is the dominant part of magnetic susceptibility above 35 K. The Van Vleck paramagnetic contribution reported by Onodera et al. \cite{Onodera1998} is in our case well modeled by the temperature-independent term $\chi_0$.

\begin{figure}[h]
 \includegraphics[scale=0.24]{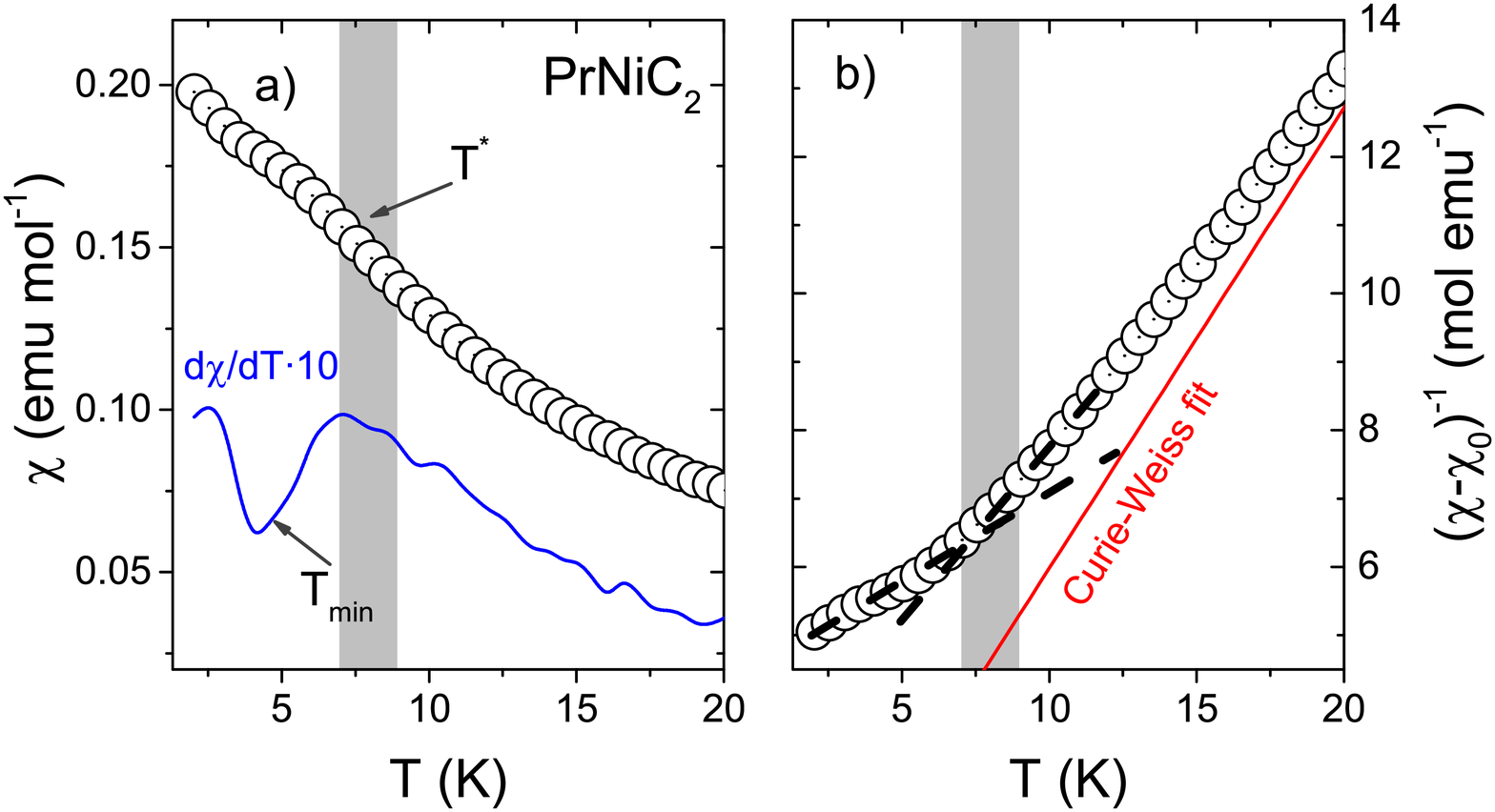}
  \caption{\label{PrAC} a) Low-temperature dc magnetic susceptibility of PrNiC$_2$ measured at 1 T applied field showing a slight upturn arround 7 K, below the magnetic anomaly temperature $T^{*}$ (see text). The differential of the dc susceptibility (blue line) shows a minimum arround 4 K. b) Inverse magnetic susceptiblity of PrNiC$_2$ corrected for the temperature independent contributions $\chi_0$. Red line shows the Curie-Weiss fit from Fig. \ref{MvsT_all} b). Dashed lines are a guide for the eye. }
  \end{figure}

Upon crossing the N\'eel temperature $T_N$ = 17 K, the magnetic susceptibility of NdNiC$_2$ drops rapidly. A similar drop, yet much less pronounced, is seen also in CeNiC$_2$ below $T_N$ = 19 K. The susceptibility of PrNiC$_2$ shows no clear sign of a magnetic transition above 2 K, in agreement with previous reports \cite{Kotsanidis1989,Onodera1998}, however a small kink in the curve is seen at $T^* \approx$ 8 K (see Fig. \ref{PrAC}), consistent with the decrease in magnetization along the \textit{a} crystallographic axis seen at this temperature by Onodera et al. \cite{Onodera1998}). The underlying cause for this magnetization anomaly is not clear, but may suggest some type of electronic or crystal structure transition, resulting in the decrease of Pauli or Van Vleck paramagnetic susceptibility.

Magnetization vs. applied field ($M(H)$) for CeNiC$_2$, PrNiC$_2$, and NdNiC$_2$ is presented in Figure \ref{Pr-Mdc_vs_T_PrNd_M_vs_H}. For CeNiC$_2$ (Fig. \ref{Pr-Mdc_vs_T_PrNd_M_vs_H}a) the magnetization is linear above $T_N$, with an upturn developing above approx. 4 T in the lower temperatures. Below the second transition temperature ($T_t$ = 7 K) hysteresis is observed in $M(H)$. Even at 9 T applied magnetic field, the magnetization reaches only 0.27$\mu_B$ which is ca. 13$\%$ of the expected saturation magnetization for Ce$^{3+}$ ion $gJ$ = 2.14 $\mu_B$ (where $g = \frac{4}{5}$ is the Lande $g$-factor, and $J = 4$ is the total angular momentum)\cite{Jensen1991}. The magnetization at 2 K and 9 T for CeNiC$_2$ is however approximately half of the observed saturation moment for a pure Ce metal which is only 0.6$\mu_B$\cite{Jensen1991}.

For PrNiC$_2$, $M(H)$ is roughly linear up to 9 T applied field at temperatures above 40 K (see Fig. \ref{Pr-Mdc_vs_T_PrNd_M_vs_H}b), below which the curves start to slightly deviate from linearity. At 10 K and below (Inset of Fig. \ref{Pr-Mdc_vs_T_PrNd_M_vs_H}b) the deviation is more pronounced and the curves start to saturate. At 2 K and 9 T applied field the $M(H)$ of PrNiC$_2$ reach approx. 1.5 $\mu_B$, which is half of the expected saturation magnetization for Pr$^{3+}$ ion $gJ$ = 3.20 $\mu_B$\cite{Jensen1991}.

In case of NdNiC$_2$, the magnetization curves are linear down to 20 K (Fig. \ref{Pr-Mdc_vs_T_PrNd_M_vs_H}c and d). Below the $T_N$ the ($M(H)$) is strongly suppressed, but above 4 T a sudden upturn is observed, resulting from field-induced magnetic order-order transition that reduces the AFM compensation of local moments. Similar transitions have been previously observed in GdNiC$_2$ \cite{Kolincio20161}. Above the transition the $M(H)$ curves start to saturate, reaching 1.6$\mu_B$ in 9 T at 2 K, about one half the saturation magnetization for Gd ion ($gJ$ = 3.27 $\mu_B$ \cite{Jensen1991}). The magnetization loop shows no trace of hysteresis at the AFM-FM transition as it is presented in the inset of Fig. \ref{Pr-Mdc_vs_T_PrNd_M_vs_H}c. 

\begin{figure}[h!t]
 \includegraphics[scale=0.25]{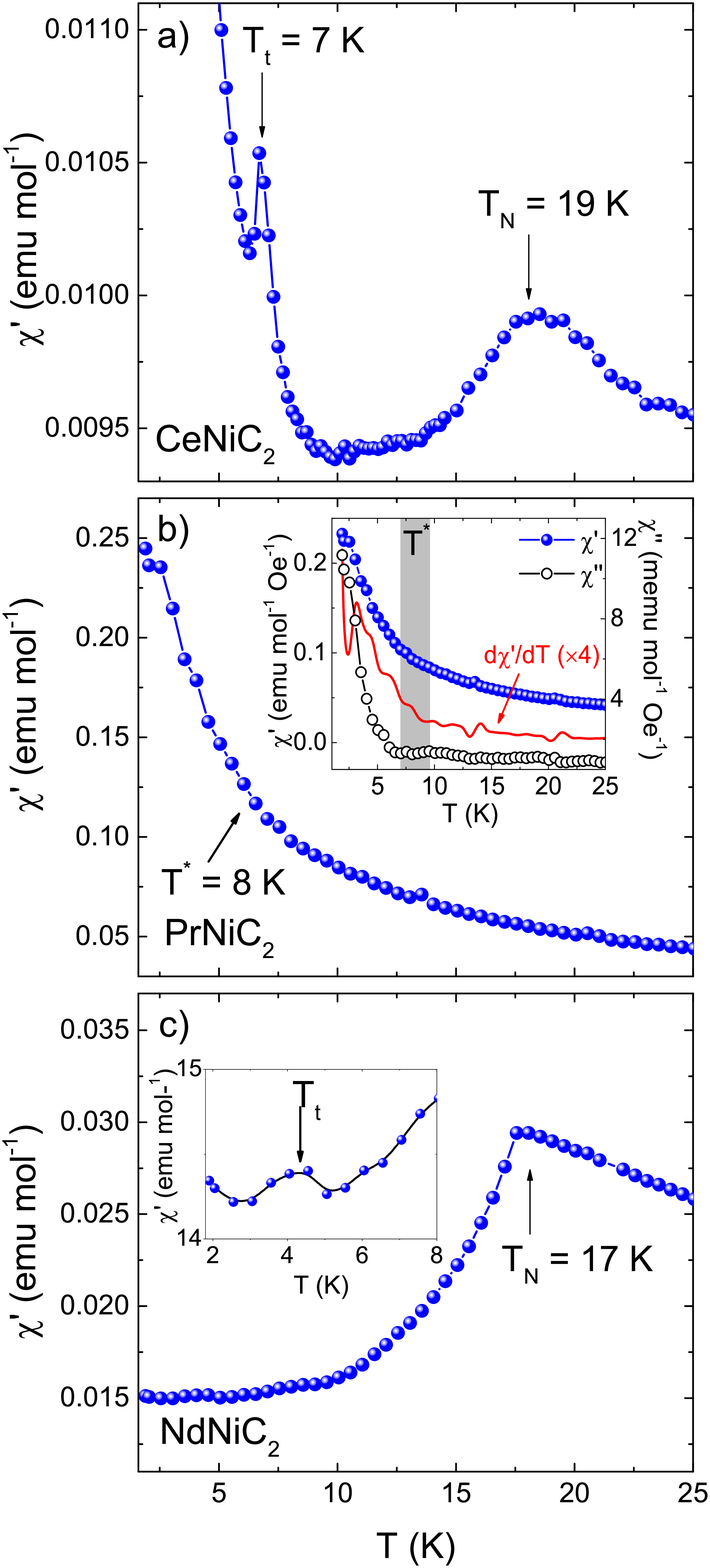}
  \caption{\label{Mac_vs_T} Real part of ac magnetic susceptibility of a) CeNiC$_2$, b) PrNiC$_2$, c) NdNiC$_2$ measured in a constant field of 5 Oe with 3 Oe, 1 kHz excitations. Blue arrows on panel a indicate the transition to an AFM state at $T_N$ = 19 K and order-order transition at approx. 7 K.  Inset of panel b presents the comparison of real and imaginary parts of the ac susceptibility (blue and black points, respectively) and the derivative of the real part (red line). The value of derivative is negative and decreases with decreasing temperature. In panel c the $T_N$ = 17 K is defined as a position of the drop of susceptibility at the AFM transition. Inset shows a small jump around 4 K that is attributed to
magnetic order-order transition.}
  \end{figure}
  
The real part of the ac magnetic susceptibility of CeNiC$_2$ and NdNiC$_2$ shows a drop at the N\'eel temperature $T_N$ of 19 and 17 K, respectively (see Fig. \ref{Mac_vs_T}a,c), in agreement with previous reports \cite{Onodera1998}. Below $T_N$ both compounds undergo further magnetic transitions. In CeNiC$_2$ a sudden drop of susceptibility is seen at $T_t$ = 7 K followed by a pronounced upturn. The change in magnetic order below 10 K was previously observed by magnetization, specific heat and NMR measurements\cite{Motoya1997,Onodera1998}. An additional small upturn around 29 K results from the presence of a minor quantity of the antiferromagnetic CeC$_2$ impurity phase\cite{Sakai1980} ($T_N$ = 30 K), observed in XRD measurements. In NdNiC$_2$ a small feature is seen around 4 K (see the inset of Fig. \ref{Mac_vs_T}c) that was reported by Onodera \textit{et al.}\cite{Onodera1998}. The ac susceptibility of PrNiC$_2$ shows no clear sign of magnetic transition, however the slightly saturating dependency of $\chi'$ and its derivative $d\chi'/dT$ resembles the results obtained for the Pb$_2$Sr$_2$PrCu$_3$O$_8$ compound in which a quasi-2D magnetic order is observed below 7 K as evidenced by neutron diffraction study \cite{Hsieh1994}. In the aforementioned case the ac susceptibility show a saturation below the ordering temperature rather than a pronounced drop while the differential exhibit a minimum at the ordering temperature. In our case there is no clear minimum of the differential curve, yet it would be necessary to perform a neutron diffraction measurement in order to confirm or deny the presence of long-range magnetic order below the $T^*$.

In contrast with CeNiC$_2$ and NdNiC$_2$, PrNiC$_2$ does not reveal any clear magnetic transition. Since the three compounds are chemically similar, the discrepancy arises likely from the difference in the detailed structure of 4\textit{f} energy levels. The ground state of a free Pr$^{3+}$ ion is ninefold degenerate with total angular momentum $J$ = 4. The crystalline electric field (CEF) acting on the Pr$^{3+}$ removes the degeneracy (either fully or partially), with the nature of the effect dependent on the point symmetry of the ion crystallographic position. In the orthorhombic PrNiC$_2$ the $2a$ site occupied by a Pr atom has the point symmetry group $mm2$. For such relatively low symmetry one would expect a complete uplifting of the ground state degeneracy, yielding a nonmagnetic configuration with 9 separated singlet states similarly as in PrNi$_2$Al$_5$\cite{Akamaru2001}. Note however that in the case of exchange interaction energy exceeding the first CEF excitation, the magnetic order may appear due to the intermixing of higher energy states into a ground state with higher degeneracy\cite{Snyman2013}. Such situation occurs in the orthorhombic PrNiGe$_2$ compound crystallizing in the CeNiSi$_2$-type structure (related to CeNiC$_2$) in which the Pr$^{3+}$ ion position has the same point symmetry as in PrNiC$_2$, yet the material reveals ferromagnetic (FM) ordering at $T_C$ = 13 K\cite{Gil1994,Snyman2013}.

\begin{figure}[ht]
 \includegraphics[scale=0.24]{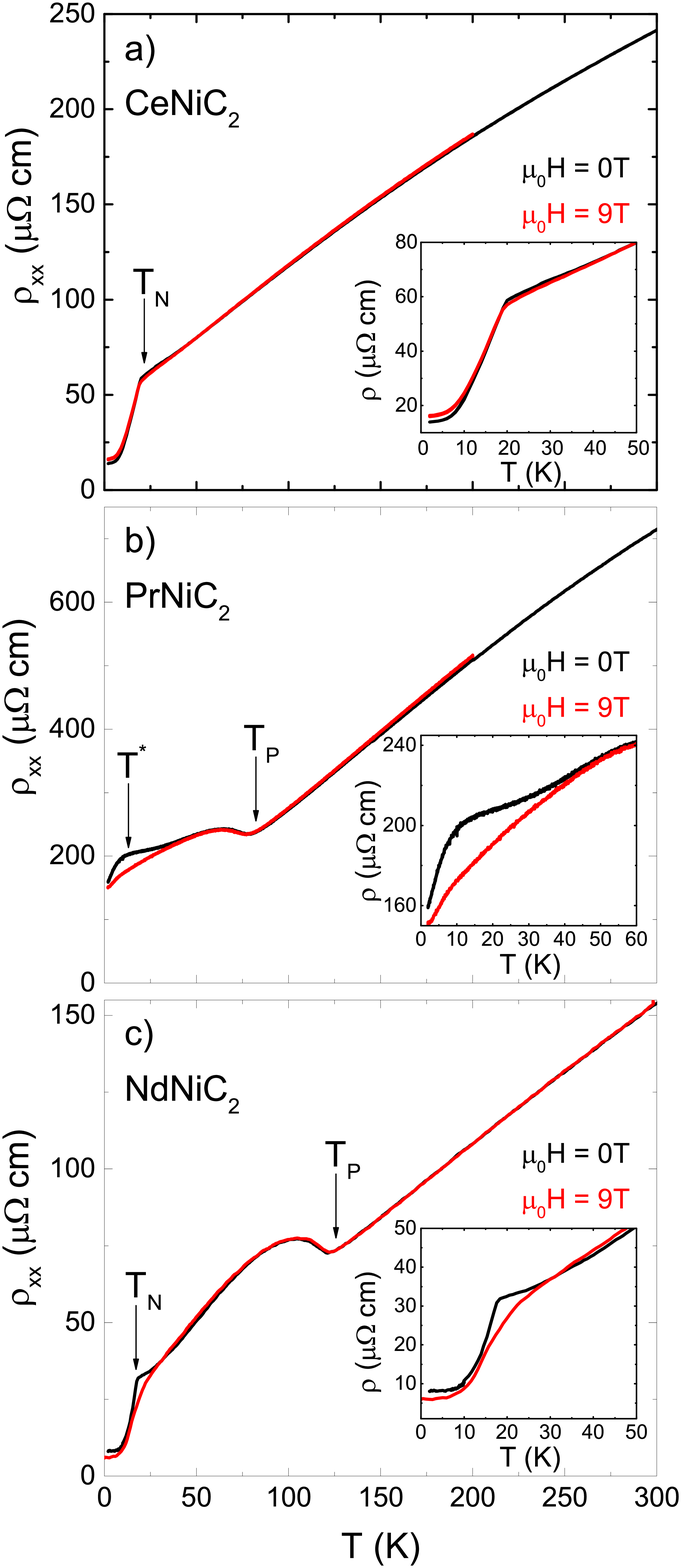}
  \caption{\label{rho} Resistivity of a) CeNiC$_2$, b) PrNiC$_2$, c) NdNiC$_2$,  measured without (black color) and with (red color) applied magnetic field of 9 T. Arrows indicate characteristic temperatures: $T_P$ - Peierls temperature for NdNiC$_2$ and PrNiC$_2$, $T_N$ N\'eel temperature for CeNiC$_2$ and NdNiC$_2$, and $T^*$ - magnetic anomaly temperature in PrNiC$_2$. Insets: Expanded view of the vicinity of the magnetic ordering (anomaly) temperature.}
  \end{figure}

Figure \ref{rho}a, b and c, shows the thermal dependencies of electrical resistivity ($\rho_{xx}$) measured without and with applied magnetic field (9 T), for CeNiC$_2$, PrNiC$_2$ and NdNiC$_2$ respectively. At high temperatures, all the compounds exhibit typical metallic behavior with resistivity deceasing with  temperature lowering. Upon cooling, $\rho_{xx}$ of both PrNiC$_2$ and NdNiC$_2$ show the anomalies pronounced by a minimum followed by a hump. This metal-metal transition is a typical signature of the charge density wave state with incomplete Fermi surface nesting, characteristic for quasi-2D materials \cite{Kolincio20162}. The temperature of this anomaly corresponds to the Peierls temperature ($T_P$ = 121 K for NdNiC$_2$ and $T_P$ = 89 K for PrNiC$_2$) established by X-ray diffuse scattering\cite{Yamamoto2013}. In contrast to that, no CDW-like anomaly is observed in the third compound, CeNiC$_2$.  At the magnetic crossover temperatures, all three curves exhibit a decrease in resistivity, shown closer in the insets of Figure  \ref{rho}. This downturn is visibly sharper for the antiferromagnetic ground states of NdNiC$_2$ and CeNiC$_2$ than in the case of PrNiC$_2$, where instead of a long range of magnetic ordering, one observes a small magnetic anomaly at $T^*$.

Although the anomalies in the zero field resistivity have been reported beforehand\cite{murase2004}, the influence of magnetic field on transport properties, up to now, has been studied solely for the Nd-bearing compound\cite{Yamamoto2013, Lei2017}. Electrical resistivity measured in the presence of a magnetic field of $\mu_0H$ = 9 T is shown as a red line in Figure \ref{rho}, a b and c. The influence of magnetic field on $\rho_{xx}$ in the high temperature metallic state of each compound is negligibly small. In CeNiC$_2$, this behavior is present down to the vicinity of $T_N$, where the magnetic field weakly modifies the resistivity. This is in contrast to the features seen in the two compounds exhibiting the charge density waves; in NdNiC$_2$ one observes a notable decrease in resistance with magnetic field at $T\rightarrow T_N$. In PrNiC$_2$ the onset of the negative magnetoresistance can be observed at $T\approx$ 60 K, much closer to $T_P$ than in NdNiC$_2$.
To investigate further the impact of $\mu_0H$ on transport properties of studied compounds we have performed the field sweeps at constant temperatures. 

\begin{figure}[t]
 \includegraphics[angle=0,width=0.8\columnwidth]{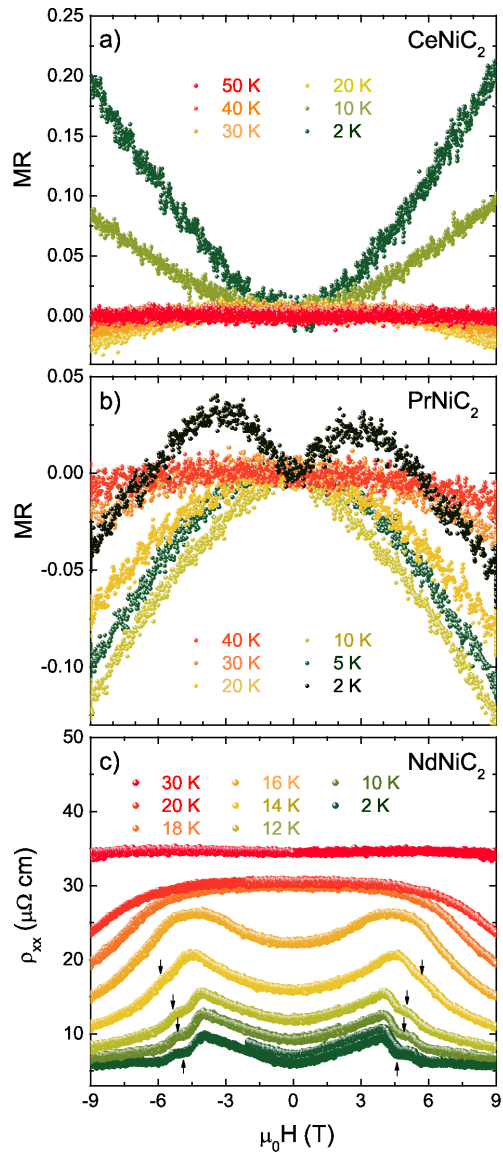}
  \caption{\label{MR} Magnetotransport properties of $R$NiC$_2$. All the measurements have been performed at constant temperature. a) Magnetoresistance in CeNiC$_2$ as a function of magnetic field, b) Magnetic field dependence of magnetoresistance in PrNiC$_2$, c) Resistivity of NdNiC$_2$ as a function of magnetic field. For better clarity, for this compound we show the $\rho_{xx}$ instead of MR. Arrows indicate the kinks attributed to a metamagnetic phase separating the FM and AFM orders.}
  \end{figure}

The magnetic field dependence of magnetoresistance (MR = $\frac{\rho(H)-\rho_0}{\rho_0}$, where $\rho_0$ is the zero field resistivity) of CeNiC$_2$ is depicted in Figure \ref{MR}a. At $T > T_N$, MR is weak and negative (resistivity decreases by a maximum of 3\%). Below this temperature, the magnetoresistance changes its sign and magnitude. This is a typical picture of the modification of the scattering rate in the vicinity of the magnetic ordering temperature\cite{Usami1978, Mazumdar1997, Yamada1972}; above $T_N$ the reduction of resistance can be attributed to the field induced ordering of the local magnetic moments, resulting in the quenching of the spin fluctuations and effectively a decrease of the related scattering mechanism. On the other side of the transition, below $T_N$, the magnetic field induces a partial reorientation of the local spins and perturbs the antiferromagnetic order, which results in the increase of the scattering rate and, consequently, of the electrical resistance.

   Figure \ref{MR}b shows the magnetic field dependence of magnetoresistance of PrNiC$_2$. One can notice that, in the charge density wave state, MR is dominated by the negative component which rises as temperature decreases down to $T^*$. Below this temperature limit, the negative MR decreases and finally at $T$ = 2 K a positive term can be observed at low magnetic field. This positive MR component can originate from an onset of another magnetic-like transition at lower temperatures or from the light carriers related to the small Fermi surface pockets that can be opened in the FS due to imperfect nesting. A complementary experiment,  such as ARPES spectroscopy, neutron diffraction or  magnetotransport measurements performed at temperatures below 1.9 K and higher field would be required to clarify this point.
  Figure \ref{MR}c shows the magnetic field dependence of resistivity of NdNiC$_2$. Due to the rich variety of positive and negative MR components seen in this compound, we find it more clear to use the $\rho_{xx}(H)$ instead of MR(H) for discussion of the magnetotransport properties in NdNiC$_2$. At 30 K, one observes an onset of the negative magnetoresistance term, which becomes stronger as temperature decreases. Below $T_N$, the resistivity firstly rises with magnetic field and after reaching the maximum, the $\rho_{xx}$ decreases again. The position of the resistivity maximum at various temperatures below $T_N$ corresponds to the magnetic field induced ferromagnetic transition according to the $H$-$T$ phase diagram of NdNiC$_2$ constructed for a single crystal \cite{Onodera1998}. Below 14 K, one observes an additional kink (marked in Fig. \ref{MR} by arrows) on the decreasing side of resistance. This can be attributed to the intermediate magnetic phase separating the AFM and FM orders at this temperature range. In addition, one can notice that at the lowest temperatures the resistivity saturates at high magnetic fields. 
The negative magnetoresistance in NdNiC$_2$ has been attributed\cite{Yamamoto2013, Lei2017} both to the suppression of spin disorder scattering and to the destruction of the charge density wave as seen in the isostructural, albeit ferromagnetic compound, SmNiC$_2$ in which the relevance of the CDW suppression has been confirmed by the X-ray diffuse scattering experiment performed in magnetic field\cite{Shimomura2009, Hanasaki2012}. 

\begin{figure*}[ht]
   \includegraphics[angle=0,width=1.8\columnwidth]{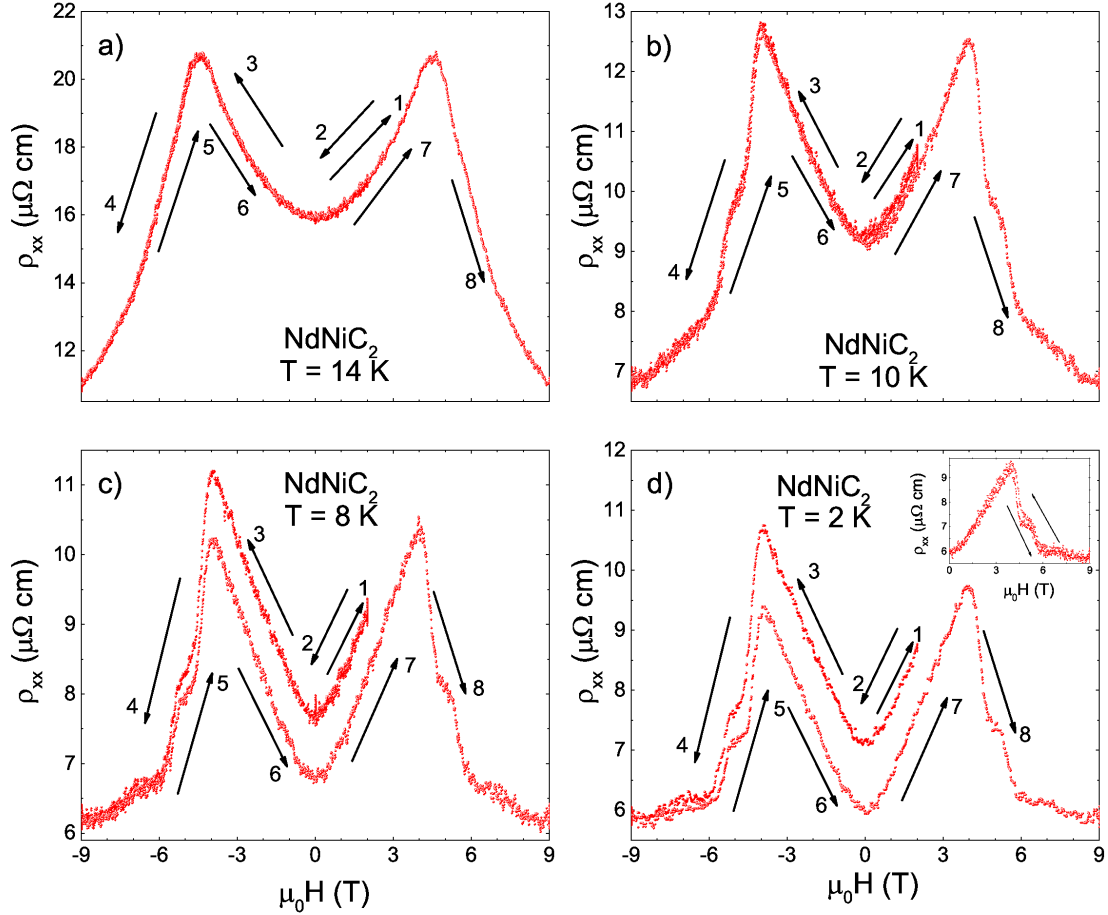}
  \caption{\label{Nd_MR_hist} Resistivity of NdNiC$_2$ measured at selected temperatures. After each field sweep data collection at constant temperature, the sample was warmed up to 40 K in zero magnetic field to remove the magnetic memory of the material. Arrows and numbers show the direction of field sweeps. a) $T$ = 14 K, b) $T$ = 10 K, c) $T$ = 8 K, d) $T$ = 2 K. Inset: Resistivity  at $T$ = 2 K of the same sample of NdNiC$_2$, however previously subjected to the magnetic field of 9 T at $T$ = 5 K.}
  \end{figure*}

  An interesting observation is the irreversible behavior of the electrical resistivity at low temperatures. In order to prove that this effect is not an artifact caused by unstable electrical contacts and is intrinsic to the sample, we have repeated the measurement at lower temperatures. Firstly the sample was warmed up to 40 K, far above the magnetic ordering temperature ($T_N=$ 17 K). Next, we have cooled the sample with zero applied field, and stabilized the temperature before activating the magnet. The magnetic field was swept initially to 2 T, to avoid crossing the AFM-FM transition. Then, the magnetic field was swept and reached -9 T (9 T applied in the adverse direction). Afterwards, we performed the final sweep and continuously reversed the direction of the magnetic field to 9 T. The whole procedure was repeated for each scan in order to remove any magnetic memory from the sample. In Figure \ref{Nd_MR_hist} we show the results of the field sweeps at the selected temperatures.
The resistivity measured at $T$ = 14 K (Figure \ref{Nd_MR_hist}a) is reversible with $\mu_0H$. At $T$ = 10 K (Figure \ref{Nd_MR_hist}b) one can notice a small irreverisibility of $\rho_{xx}$, which becomes more pronounced at $T$ = 8 K, as depicted in Figure \ref{Nd_MR_hist}c. When the magnetic field is increased to 2 T and then swept to 0, the resistivity returns to the zero-field cooled value of $\rho_0$. In these conditions, the sample remains in the AFM state. However, the application of a magnetic field exceeding the limit of 4 T, at which the FM order is induced in the sample, prevents the resistance from returning to the original $\rho_0$. Further magnetic field sweeps do not induce any irreversible transitions and the resistivity returns to the new value of $\rho_0^*$ when the field is reduced back to 0. Figure \ref{Nd_MR_hist}d compares the result of a field sweep of the sample cooled to 2 K in ZFC condition and the $\rho_{xx}$ of the same sample, which previously experienced the transformation to the FM state at $T$ = 5 K (inset). The irreversible behavior is clearly visible in the former case, while in the latter one the resistivity returns to the initial value.  This shows that the resistance of NdNiC$_2$ depends not only on temperature, applied magnetic field or the type of  magnetic ordering present in the sample at these conditions, but also on the magnetic history of the sample and this metastable effect is clearly associated with the AFM-FM transition. Previous reports on the magnetoresistance of NdNiC$_2$ \cite{Yamamoto2013, Lei2017} have not mentioned the irreversible phase transition, probably because this weak crossover could be easily overlooked, since once the sample experiences the high magnetic field at temperature below 12 K it remains in the metastable state and the irreversibility is no longer observable until the sample is reheated and cooled down again. One plausible scenario to explain this irreversible effect is the magnetoplastic lattice deformation induced by the ferromagnetic transition. Note that even a small lattice transformation and a consequent Fermi surface modification can substantially impact the nesting conditions and this can lead to the quasi-permanent suppression of CDW.


The BCS approach predicts the negative magnetoresistance in CDW systems to originate from the Zeeman splitting of the conduction bands\cite{Dieterich1973} which results in reduction of the pairing interactions and degradation of nesting properties. This term has been found to originate both from orbital effects and from local spins producing stronger magnetic moments. For magnetic fields $\mu_BH\ll\Delta_{CDW}$, the Zeeman magnetoresistance term is expressed\cite{Tiedje1975} by Equation \ref{Zeeman}:

\begin{equation}
\label{Zeeman}
MR = \frac{\rho (H)-\rho_{0}}{\rho_0}= -\frac{1}{2} \left( \frac{\mu_BH}{k_BT}\right)^2+0 \left( \frac{\mu_BH}{k_BT}\right)^4
\end{equation}

\begin{figure}[h]
 \includegraphics[angle=0,width=0.8\columnwidth]{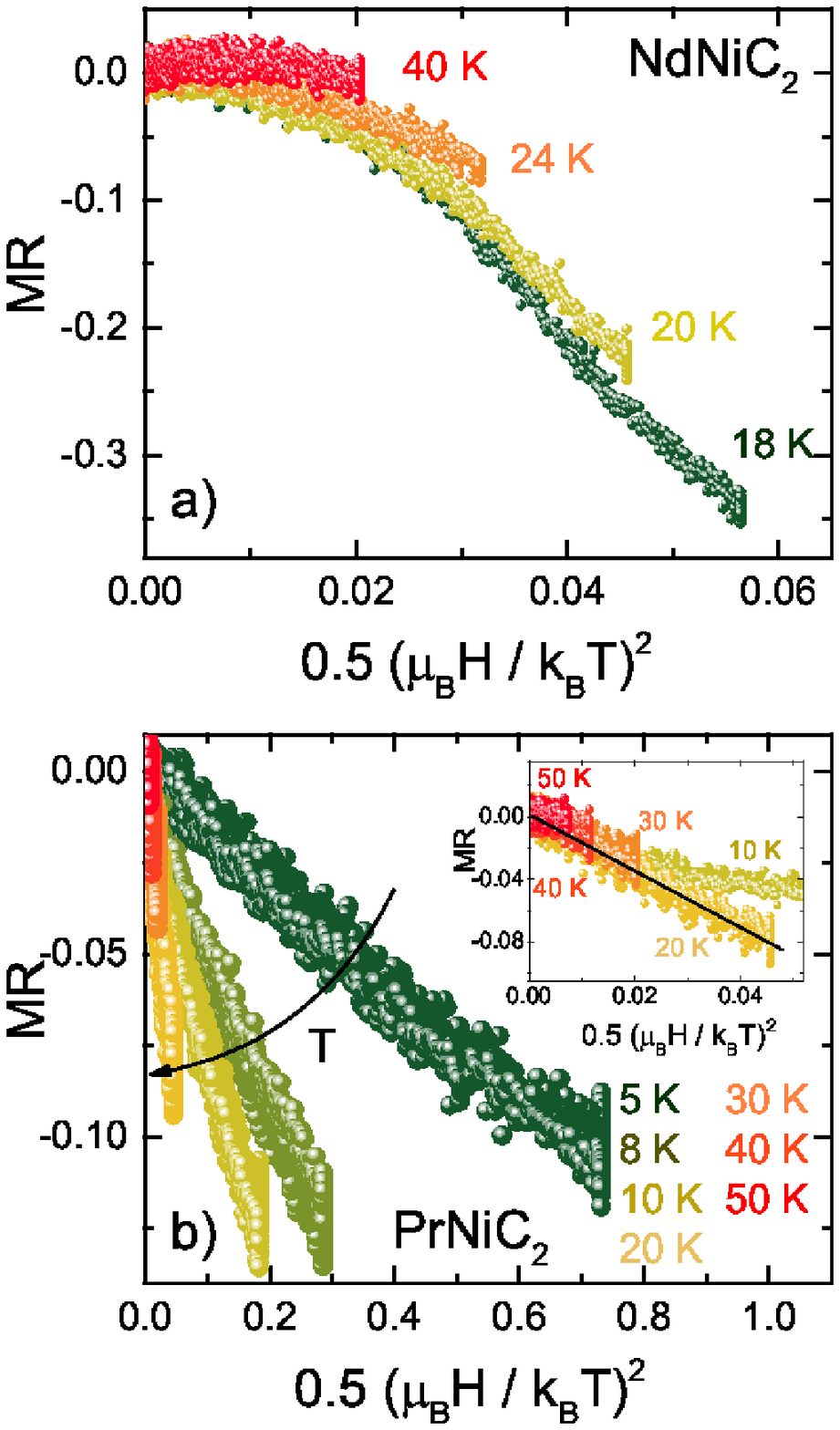}
  \caption{\label{MR_scalling} Scaling of magnetoresistance in PrNiC$_2$ wit Equation \ref{Zeeman}. Inset: Expanded view the MR scaling for $T$ $\geq$ 10 K }
  \end{figure}

The Figure \ref{MR_scalling}a shows the magnetoresistance of NdNiC$_2$ above $T_N$ as a function of $\frac{1}{2} \left( \frac{\mu_BH}{k_BT}\right)^2$. The plots do not converge into a single straight line. This is not surprising, since this temperature interval corresponds to the onset of the field induced magnetic ordering. This can lead either to the previously suggested CDW suppression, stronger than predicted by Equation \ref{Zeeman} or to the reduction of the spin scattering, which also results in negative magnetoresistance as in CeNiC$_2$. The comparison of the strength of the negative magnetoresistance in NdNiC$_2$ and CeNiC$_2$ in the vicinity of $T_N$ can also be a useful guide. In the former compound, showing the Peierls instability, MR reaches -40 \% which is an order of magnitude larger than in the latter one, in which the CDW is absent. This suggests that, the negative magnetoresistance in NdNiC$_2$  originates, at least partially, from the suppression of the CDW state.

The negative MR in PrNiC$_2$ reaches a maximum of 12\%, which although is visibly weaker than in NdNiC$_2$, still exceeds the value found in CeNiC$_2$. This, similar to the case of NdNiC$_2$, suggests that the decrease of resistance in magnetic field originates from the suppression of the CDW. To verify this hypothesis, we have scaled the magnetoresistance in PrNiC$_2$ with Equation \ref{Zeeman}, as shown in Figure \ref{MR_scalling} b. At $T>$ 20 K the PrNiC$_2$ can be qualitatively described by the Zeeman term; the MR plots fall into a single straight line. At lower temperatures, in the vicinity of $T_M$ the negative magnetoresistance is weakened and diverges from this scalling law (as shown in the inset of Figure \ref{MR_scalling}b). The  curve obtained for $T$ =  10 K is a boundary of the relevance of the Equation \ref{Zeeman}. At $\frac{1}{2} \left( \frac{\mu_BH}{k_BT}\right)^2\approx$ 0.02, which corresponds to $\mu_B H$ = 6 T at this temperature, the magnetoresistance plot diverges from the Zeeman scaling and starts decreasing. We find that, to apply Equation \ref{Zeeman} one has to use the prefactor of approximately 1.4. In other CDW materials this coefficient is usually smaller than unity. The key examples are Li$_{0.9}$Mo$_6$O$_{17}$ \cite{Xu2009} or organic compounds such as (Per)$_2$Pt(mnt)$_2$ \cite{Graf20041, Matos1996, Bonfait1995, Bonfait1991} in which the existence of weakly magnetic chains ramps this magnetoresistance prefactor in comparison with (Per)$_2$Au(mnt)$_2$ \cite{Graf2005, Monchi1999} showing a non-magnetic character. On the other hand, the value we found is  significantly lower than the factor of $\approx$ 30 found in GdNiC$_2$ \cite{Kolincio20161}, where the presence of strong local magnetic moments amplifies the internal magnetic field much more effectively than in PrNiC$_2$, showing no clear long range magnetic ordering.  

\begin{figure}[h]
 \includegraphics[scale=0.4]{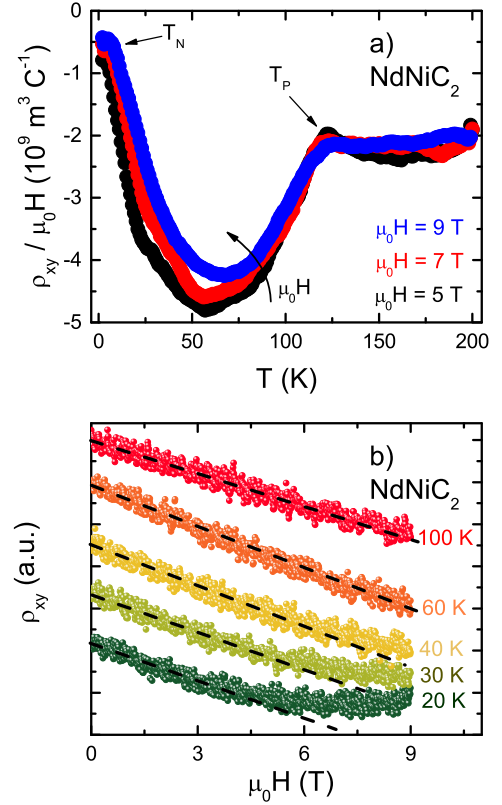}
  \caption{\label{Nd_Hall} a) Hall resistivity of NdNiC$_2$, divided by magnetic field, measured at various magnetic fields. Arrows indicate the Peierls and N\'eel temperatures $T_P$ and $T_N$ respectively. b) Hall resistivity of NdNiC$_2$ as a function of magnetic field. The plots have been shifted horizontally to improve data reading. }
  \end{figure}

Due to polycrystalline nature of our samples, we are unable to perform the X-ray diffuse scattering experiment to follow the intensity and position of the satellite reflections at various temperature and magnetic field. Instead, to investigate the suppression of the charge density waves state by magnetic field, we have conducted the Hall effect measurements, which can be used as a direct probe for electronic carrier concentration. 
Figure \ref{Nd_Hall}a shows the thermal dependence of Hall resistivity ($\rho_{xy}$) in NdNiC$_2$.  The sign of the measured Hall resistance is negative, opposite to the results reported recently \cite{Lei2017}. To clarify this point, we have repeated the measurement with a reference sample of Cu foil, which shows a negative Hall signal in the same contact geometry. This confirms the relevance of the negative sign of $\rho_{xy}$ in NdNiC$_2$. At $T>T_P$, the Hall signal is almost independent of temperature. At the Peierls temperature one observes a downturn of $\rho_{xy}(T)$ (and increase of $\vert\rho_{xy}\vert$), which is a typical signature of the opening of the CDW bandgap and condensation of electronic carriers \cite{Wang1989, Schlenker1985}. Upon further cooling, the Hall resistivity decreases until it reaches a minimum followed by a prominent increase of $\rho_{xy}$ (and decrease of $\vert\rho_{xy}\vert$), which grows even higher than for temperatures above $T_P$.


This increase of $\rho_{xy}$ in proximity of the magnetic ordering temperature observed in SmNiC$_2$\cite{Kim2012} and NdNiC$_2$\cite{Lei2017} has been attributed to the destruction of CDW and a concomitant release of previously condensed carriers. Although the CDW suppression by magnetic field appears to be quite a possible scenario, this mechanism itself is not sufficient to explain the features observed as $T \rightarrow T_N$, especially considering that the low temperature $\vert\rho_{xy}\vert$ is lower than the value found for $T>T_P$. This could lead to a misguiding suggestion that the carrier concentration below $T_N$ exceeds the high temperature normal state value. To avoid the oversimplification, in a material exhibiting magnetic ordering, one has to consider two components of the Hall resistance\cite{Berger1980}:

\begin{equation}
\label{Hall_eq}
\rho_{xy}=R_0\mu_0H+4\pi R_SM
\end{equation}

The $R_0$ in Equation \ref{Hall_eq} is the ordinary Hall coefficient which, in a single band model, is inversely proportional to the carrier concentration. $R_S$ denotes the anomalous Hall coefficient associated with side jump and skew scattering. To obtain the more clear evidence of the partial CDW destruction in NdNiC$_2$, we complement the previous Hall effect study\cite{Lei2017} of this compound in regard to the anomalous component of the Hall signal. We also present the results of the same experiment for CeNiC$_2$ and PrNiC$_2$ which similarly to magnetoresistance in these two compounds have not been reported previously. The separation of normal and anomalous $\rho_{xy}$ components is not straightforward unless the magnetic moment saturates with magnetic field which then reduces the latter one to a constant\cite{Higgins2004, Xu2006, Oiwa1999, Shiomi2009}. Here, no signs of saturation of $M(T)$ up to  an applied field of 14 T for any of the studied compounds have been found\cite{Rogacki}, which precludes the possibility of the direct extraction of electronic concentration from $\rho_{xy}$. Nevertheless we can propose an alternative road to follow the number of carriers condensed into the charge density wave state. The idea is to compare the field dependencies of $\rho_{xy} $ and $M$ with a special regard for the temperature region, in which magnetization follows the linear field dependency. In this condition the anomalous component contribution is also linear with field and, for a single band metal, any departure from the the linearity of $\rho_{xy}$ indicates the change of $R_0$ which is a measure of electronic concentration. 

Figure \ref{Nd_Hall}b shows the magnetic field dependence of the Hall resisitivity of NdNiC$_2$ measured at various temperatures. At $T \geq$ 60 K one cannot find any departure from linearity for the $\rho_{xy}(H)$. A small nonlinearity can be seen at 40 K. Upon further cooling, the deviation from linear variation for $\rho_{xy}(T)$ becomes more pronounced. Comparing this result with magnetization data for NdNiC$_2$ (Fig. \ref{Pr-Mdc_vs_T_PrNd_M_vs_H}d), which shows linear $M(H)$ dependence at $T \geq$ 20 K one can deduce that, in this temperature range, the non-linearity of $\rho_{xy}(H)$  can be safely attributed to the increase in electronic concentration. This indicates that, the release of previously CDW condensed carriers is, next to the anomalous Hall component, responsible for the increase of $\rho_{xy}$ as temperature is lowered to the vicinity of $T_N$. Here we emphasize that, since we were unable to observe the saturation of $M(H)$ we are unable to separate the normal and anomalous components of the Hall resistivity for $T\leq$ 20 K, where both $\rho_{xy}$ and $M$ are non-linear functions of $\mu_0H$. 
\begin{figure}[h]
 \includegraphics[scale=0.4  ]{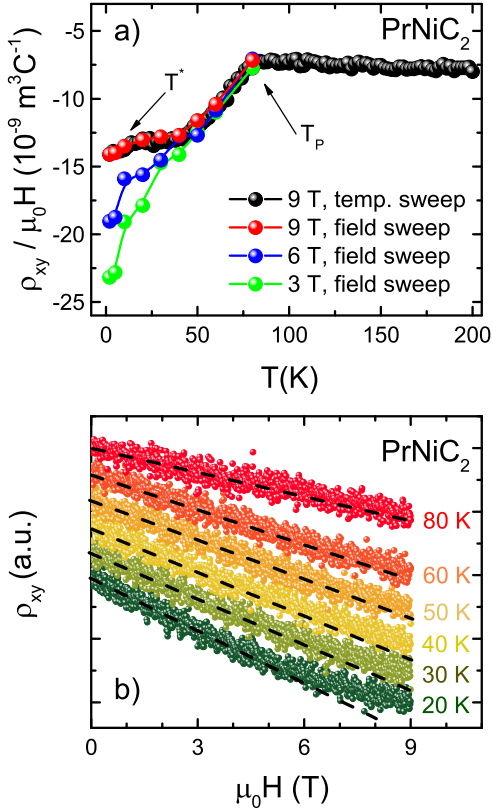}
  \caption{\label{Pr_Hall} a) Hall resistivity of PrNiC$_2$, divided by magnetic field, black points show the data collected from the temperature sweep at constant magnetic field of 9 T. Red, blue and green points show the data collected from the field sweeps at constant temperature. Arrows indicate the Peierls and magnetic transition temperatures $T_P$ and $T^*$ respectively. Solid lines are the guide for the eye. b) Hall resistivity of PrNiC$_2$ as a function of magnetic field. The plots have been shifted horizontally to improve data reading. Dashed lines show the low field linear dependencies of $\rho_{xy}(H)$ expanded to the high field regime.}
  \end{figure}
The thermal dependence of Hall resistance of PrNiC$_2$ depicted in Figure \ref{Pr_Hall}a exhibits some similarities to the case of NdNiC$_2$. A significant downturn of $\rho_{xy}$ below $T_P$ concomitant with an increase of resistivity (Figure \ref{rho}c) due to the condensation of the electronic carriers is observed at $T_P$. Upon further cooling, the Hall resistivity continues to decrease and does not simply saturate at $\frac{T_P}{2} $, where the electronic gap is expected to be fully open. This behavior is consistent with the non-BCS thermal dependence of the satellite reflections intensity\cite{Yamamoto2013} suggesting that the nesting vector adjusts to the FS evolution. In contrast to NdNiC$_2$, no significant upturn of $\rho_{xy}$ is observed as $T$ approaches the magnetic ordering temperature. Contrarily, below $T^*$ the Hall resistivity starts to decrease again. This observation is in agreement with the behavior of the intensity of the CDW satellite reflections \cite{Yamamoto2013}, which show a sudden increase upon crossing $T^*$. Below $T\approx$ 60 K, corresponding to the onset of negative magnetoresistance, the $\rho_{xy}(T)$ curves obtained at different magnetic fields do not converge. The application of stronger magnetic field drives the thermal dependence of $\rho_{xy}$ towards more positive values, in comparison to the data obtained at lower $H$. Similar to NdNiC$_2$, this can be attributed to the positive anomalous Hall component growing as the magnetization increases or to the partial suppression of the CDW and increase of the electronic concentration. It shall be noted that, the strength of the $\rho_{xy}$ downturn  below $T^*$ is sufficient to overcome the anomalous term driving the Hall resistivity towards more positive values. Note that, the strength of the anomalous Hall signal in PrNiC$_2$ is expected to parallel the scale  of NdNiC$_2$, since the values of magnetization of both compounds are comparable. To explore this effect further, we have conducted $\rho_{xy}(H)$ measurements for PrNiC$_2$. As shown in Figure \ref{Pr_Hall}b, the non-linearity of the Hall resistivity plotted versus $\mu_0H$ can be observed in this compound as well. The deviation from linearity, initially barely observable for $T$  = 50 K becomes stronger at lower temperatures. Here, however, we cannot follow the same analysis as for the case of NdNiC$_2$, due to the fact that for temperatures lower than 60 K the magnetization does not follow a linear relationship with $\mu_0H$. Therefore, the two normal and anomalous ingredients of the Hall resistivity in PrNiC$_2$ cannot be unambiguously separated. Nevertheless, the downturn of $\rho_{xy}$ at $T^*$ strongly suggests the enhancement of the CDW state, although the magnetoresistance above $T^*$ shows some signatures of the partial suppression of the Peierls instability.  This can be explained in terms of the lattice transformation accompanying the magnetic anomaly modifying the Fermi surface, which triggers the nesting of another FS part when the CDW vector adjusts to band structure evolution. One cannot however exclude an alternative scenario, in which the enhancement of the Fermi surface nesting can be seen as a driving force for the magnetic anomaly. Since the magnetic properties are related to the free electron density via RKKY interactions, it is not unreasonable to expect the condensation of the electronic carriers at $T^*$  to modify of the magnetic character of PrNiC$_2$. The high resolution X-ray and neutron diffraction experiment performed with a single crystal of PrNiC$_2$ will be required to clarify this point.

\begin{figure}[t]
 \includegraphics[scale=0.3]{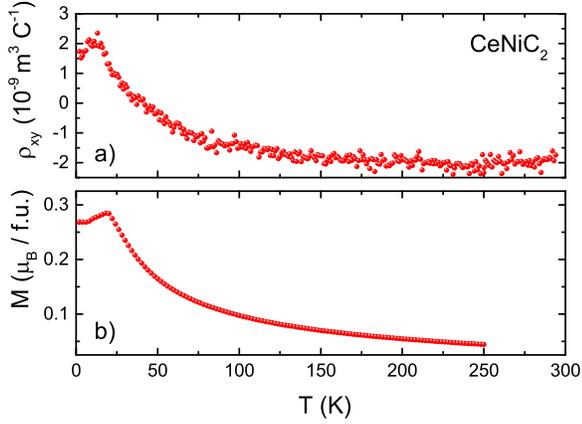}
  \caption{\label{Ce_Hall} Hall resistivity in CeNiC$_2$  as a function of temperature (a) compared with magnetization (b) of the same compound}
  \end{figure}

The thermal dependence of Hall resistivity in CeNiC$_2$, shown in Figure \ref{Ce_Hall}a shows no signatures of electronic condensation. This is in agreement with transport properties in which no anomalies similar to those found in NdNiC$_2$ and PrNiC$_2$ are observed and confirms the absence of the Peierls instability in CeNiC$_2$. From the clear correlation between the thermal dependence of $\rho_{xy}$ and magnetization (see Figure \ref{Ce_Hall}b), one can conclude, that the anomalous component is the dominant ingredient of the Hall effect in this compound, while the normal Hall coefficient is expected to remain temperature independent. The observation of the increase of $\rho_{xy}$ as $T\rightarrow T_N$ in CeNiC$_2$, where the absence of the CDW has been emphasized, implies that the anomalous Hall component is essential to describe the $\rho_{xy}$ in NdNiC$_2$ and PrNiC$_2$.

 \begin{figure}[h!]
 \includegraphics[scale=0.25]{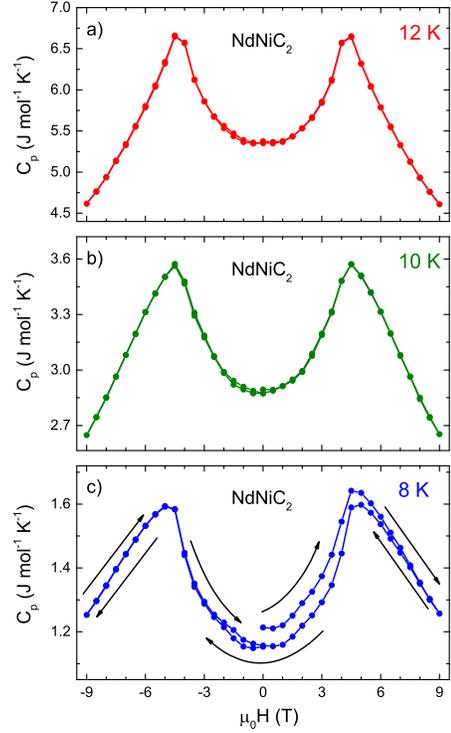}
  \caption{\label{Nd_Cp_hist} Specific heat of  NdNiC$_2$ as a function of magnetic field measured at a) $T$ = 12 K, b) $T$ = 10 K, and c) $T$ = 8 K. Arrows and numbers show the direction of the magnetic field sweeps. At each temperature step the sample was first heated to 40 K, well above the magnetic transition temperature $T_N$ = 17 K, held for a few minutes and then cooled to the target temperature with no applied magnetic field. After stabilizing the temperature, the magnetic field was first increased to 9 T, then decreased to -9 T and swept to 0 T. At 8 K an irreversible behavior is clearly seen - during the first field sweep the specific heat below 4.5 T is higher than for the second sweep from +9 to -9 T, indicating the formation of a field-induced metastable phase, which is also observed in transport measurements.}
  \end{figure}

\begin{figure*}[ht]
 \includegraphics[angle=0,width=1.8\columnwidth]{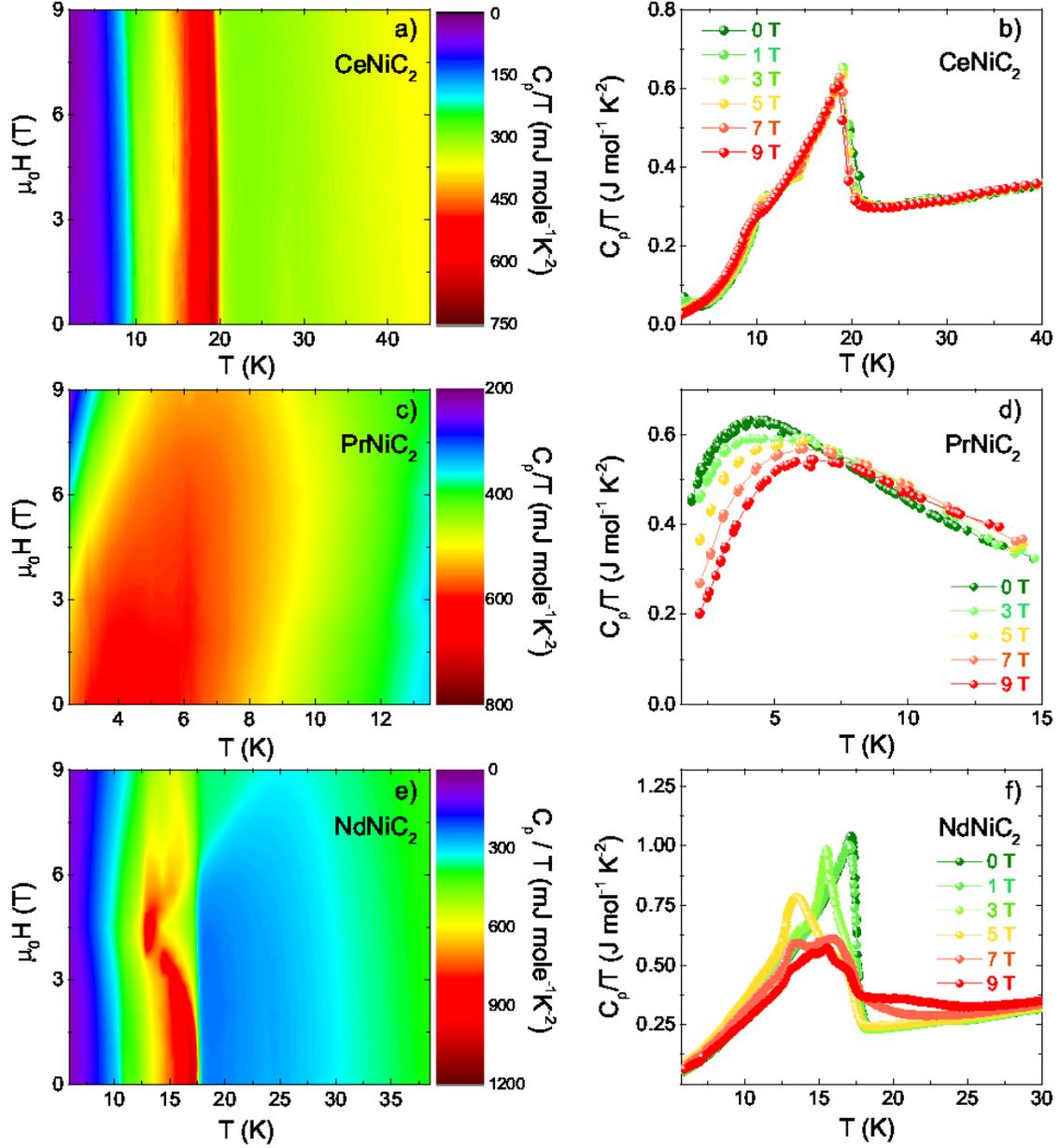}
  \caption{\label{mapa} Panels a) and b) present the specific heat of CeNiC$_2$ as a function of temperature and magnetic field. The anomaly seen at $T_N$ = 19 K does not significantly shift with applied magnetic fields up to 9 T, while the anomalies around 10 and 2 K are suppressed by increasing $\mu_0H$. Panels c) and d) show the specific heat of PrNiC$_2$, revealing that the broad hump, attributed to the Schottky anomaly resulting from splitting of the \textit{f} orbital energy levels is gradually shifted towards higher temperatures by application of a magnetic field due to the Zeeman effect. Panels e) and f) present the specific heat of NdNiC$_2$. The anomaly at 17 K remains almost unaffected by magnetic fields up to approx. 3 T above which a field-induced magnetic transition takes place, as evidenced by magnetization and transport measurements. At higher fields the specific heat curves develop a complicated structure indicating that the magnetic phase diagram is complex, as previously reported for GdNiC$_2$\cite{Kolincio20161}.}
\end{figure*}

To explore the observed transitions further, we have studied the thermal and magnetic field dependencies of specific heat ($C_p$). Previously the $C_p(T,H)$ has been successfully used to construct the phase diagram for GdNiC$_2$\cite{Kolincio20161}.
Figure \ref{mapa} shows a specific heat map (a) and the heat capacity of the polycrystalline CeNiC$_2$ (b) plotted as a function of temperature, under various magnetic fields. In the results we can observe a few anomalies. The largest one is seen at about 19 K and is almost unaffected by the applied magnetic fields up to 9 T. The second anomaly is less pronounced and the temperature of its occurrence varies with the applied magnetic field from 11 K in 0 T to 9.5 K in 9 T. The existence of the features anomalies are in agreement with magnetization and transport results. Another anomaly, previously reported by Motoya et al. \cite{Motoya1997}, seen at 2 K is magnetic field dependent. A minor jump around 30 K is likely connected with the CeC$_2$ impurity phase \cite{Sakai1980}, as suggested from magnetic susceptibility data. 

The broad hump seen in PrNiC$_2$ (Fig. \ref{mapa} c and d) is a Schottky anomaly originating from multiple energy levels of the Pr$^{3+}$ ion subject to the CEF splitting. Due to the complicated energy level structure the specific heat data could not be reliably fitted in order to extract the level splitting energies. The anomaly is slightly shifted towards higher temperature by applied magnetic field as seen in  Figure \ref{mapa} c and d, which is caused by the Zeeman effect, as seen in many \textit{f}-electron systems (see eg. \cite{Winiarski2017, Tachibana2007, Mori_2009}). No clear anomaly is seen around $T^*$ corresponding both to the drop in the Hall resistivity and the upturn of susceptibility. This may suggest that the alleged transition involves predominantly the change of electronic structure with little effect on crystal and spin order, which should result in the appearance of an anomaly in specific heat. Note that in the Pb$_2$Sr$_2$PrCu$_3$O$_8$ compound mentioned before the specific heat anomaly at the transition temperature is weak \cite{Wu1998}. If such weak anomaly would arise in PrNiC$_2$ at the $T^*$ it could be hard to observe on top of the large Schottky hump.

The results of the specific heat measurements for NdNiC$_2$ are shown in Fig. \ref{mapa} e and f. For this compound the specific heat shows a lambda-like anomaly at $T_N$, which is weakly affected by the applied magnetic field up to about 3.0-3.5 T above which a metamagnetic transition occurs. Above 7 T we can observe the third anomaly which is probably related to the occurrence of the transitional phase between AFM and FM.

The magnetic field dependence of the specific heat of NdNiC$_2$ measured at 12 K, 10 K and 8 K is presented in Fig. \ref{Nd_Cp_hist}. At 8 K the $C_p$ vs. $H$ shows an irreversible behavior as seen in Figure \ref{Nd_Cp_hist}c. The observation of the irreversibility in both specific heat and electrical resistivity measurements confirms the presence of a magnetic field-induced metastable state, not reported in previous studies. Interestingly, the same transition does not result in the appearance of hysteresis in magnetization, as seen in the inset of Figure \ref{Pr-Mdc_vs_T_PrNd_M_vs_H}. This could be explained by the insufficient resolution of magnetization measurements performed with the ACMS option. However it is also possible that the field-induced transition involves a change of electronic and crystal structures without a significant change in magnetic order.




\section{conclusions}
In order to explore the interaction between charge density waves and magnetism in the RNiC$_2$ family, we have compared the physical properties of three isostructural compounds: NdNiC$_2$, showing both the Peierls instability,  PrNiC$_2$ with the CDW and a magnetic anomaly, and CeNiC$_2$, showing antiferromagnetic ordering, and the absence of the CDW transition. 
The weak magnetoresistance in CeNiC$_2$ is found to originate by the spin fluctuations accompanying the magnetic transition. Neither transport or Hall effect measurements reveal any signatures of the Peierls instability.
Study of the magnetoresistance and the galvanomagnetic properties of NdNiC$_2$ confirms the partial suppression of charge density waves by magnetic ordering and a further destruction of the Peierls instability at the crossover from the antiferromagnetic to ferromagnetic order. We have also found that this magnetic transformation drives a metastable lattice transformation that can be observed via the magnetoresistance and the specific heat measurements. The interplay between magnetism and charge density waves in PrNiC$_2$ shows more complex character. Although the magnetoresistance data suggest that, the application of magnetic field partially suppresses CDW by Zeeman splitting of the electronic bands, the expansion of the nested region of the Fermi surface at $T^* \approx$ 8 K can be observed by a significant downturn of the Hall resistivity, strong enough to overcome the positive Hall signal originating from the anomalous component. This effect seems to be related  to the magnetic anomaly\cite{Onodera1998} observed at the same temperature, however the underlying mechanism remains unclear. Tentatively, the interaction between the CDW and magnetic properties of this compound can be described either by the lattice transformation due to the magnetic anomaly, and by the modification of the magnetic ordering via the RKKY interactions influenced by change of the electronic concentration. Further analysis of this effect can be realized by high resolution diffraction experiments on a single crystal.

\section{Acknowledgments}
Authors gratefully acknowledge the financial support from National Science Centre (Poland), grant number:  UMO-2015/19/B/ST3/03127. We also thank to E. Carnicom, K. Rogacki, Z. Sobczak, K. G\'ornicka and H. Marciniak for useful advice and fruitful discussions.
%
\end{document}